\documentclass[sigconf]{aamas} 

\usepackage{balance} 
\usepackage{booktabs}       
\usepackage{amsfonts}       
\usepackage{nicefrac}       
\usepackage{microtype}      
\usepackage{xcolor}         
\usepackage{graphicx}
\usepackage{amsmath}
\usepackage{amsfonts}
\usepackage{amsthm}
\usepackage{algorithm}
\usepackage{algorithmic}

\newtheorem{definition}{Definition}[section]

\setcopyright{ifaamas}
\acmConference[AAMAS '22]{Proc.\@ of the 21st International Conference
on Autonomous Agents and Multiagent Systems (AAMAS 2022)}{May 9--13, 2022}
{Online}{P.~Faliszewski, V.~Mascardi, C.~Pelachaud,
M.E.~Taylor (eds.)}
\copyrightyear{2022}
\acmYear{2022}
\acmDOI{}
\acmPrice{}
\acmISBN{}

\acmSubmissionID{542}

\title[AAMAS-2022 Formatting Instructions]{Multi-Agent Curricula and Emergent Implicit Signaling}

\author{Niko A. Grupen}
\affiliation{
  \institution{Cornell University}
  \city{Ithaca}
  \state{New York}
  \country{USA}}

\author{Daniel D. Lee}
\affiliation{
  \institution{Cornell Tech}
  \city{New York}
  \state{New York}
  \country{USA}}

\author{Bart Selman}
\affiliation{
  \institution{Cornell University}
  \city{Ithaca}
  \state{New York}
  \country{USA}}

\begin{abstract}
    Emergent communication has made strides towards learning communication from scratch, but has focused primarily on protocols that resemble human language. In nature, multi-agent cooperation gives rise to a wide range of communication that varies in structure and complexity. In this work, we recognize the full spectrum of communication that exists in nature and propose studying lower-level communication. Specifically, we study emergent implicit signaling in the context of decentralized multi-agent learning in difficult, sparse reward environments. However, learning to coordinate in such environments is challenging. We propose a curriculum-driven strategy that combines: (i) velocity-based environment shaping, tailored to the skill level of the multi-agent team; and (ii) a behavioral curriculum that helps agents learn successful single-agent behaviors as a precursor to learning multi-agent behaviors. Pursuit-evasion experiments show that our approach learns effective coordination, significantly outperforming sophisticated analytical and learned policies. Our method completes the pursuit-evasion task even when pursuers move at half of the evader's speed, whereas the highest-performing baseline fails at $80\%$ of the evader's speed. Moreover, we examine the use of implicit signals in coordination through position-based social influence. We show that pursuers trained with our strategy exchange more than twice as much information (in bits) than baseline methods, indicating that our method has learned, and relies heavily on, the exchange of implicit signals.
\end{abstract}

\keywords{Multi-Agent Systems, Reinforcement Learning, Communication}

\newcommand{\BibTeX}{\rm B\kern-.05em{\sc i\kern-.025em b}\kern-.08em\TeX}

\begin{document}


\pagestyle{fancy}
\fancyhead{}


\maketitle 


\section{Introduction}
Communication is a critical scaffolding for coordination. It enables humans and animals alike to coordinate on complex tasks, synchronize plans, allocate team resources, and share missing state. Understanding the process through which communication emerges has long been a goal of philosophy, linguistics, cognitive science, and AI. Recently, advances in multi-agent reinforcement learning (MARL) have propelled computational studies of emergent communication that examine the representations and social conditions necessary for communication to emerge in situated multi-agent populations \cite{lazaridou2020emergent}. Existing approaches, targeting language-like communication, have shown that it is possible for agents to learn protocols that exhibit language-like properties such as compositionality \cite{chaabouni2020compositionality,resnick2019capacity} and Zipf's Law \cite{chaabouni2019anti} when given additional learning biases \cite{eccles2019biases}. 
\begin{figure}
  \centering
  \includegraphics[width=.9\linewidth]{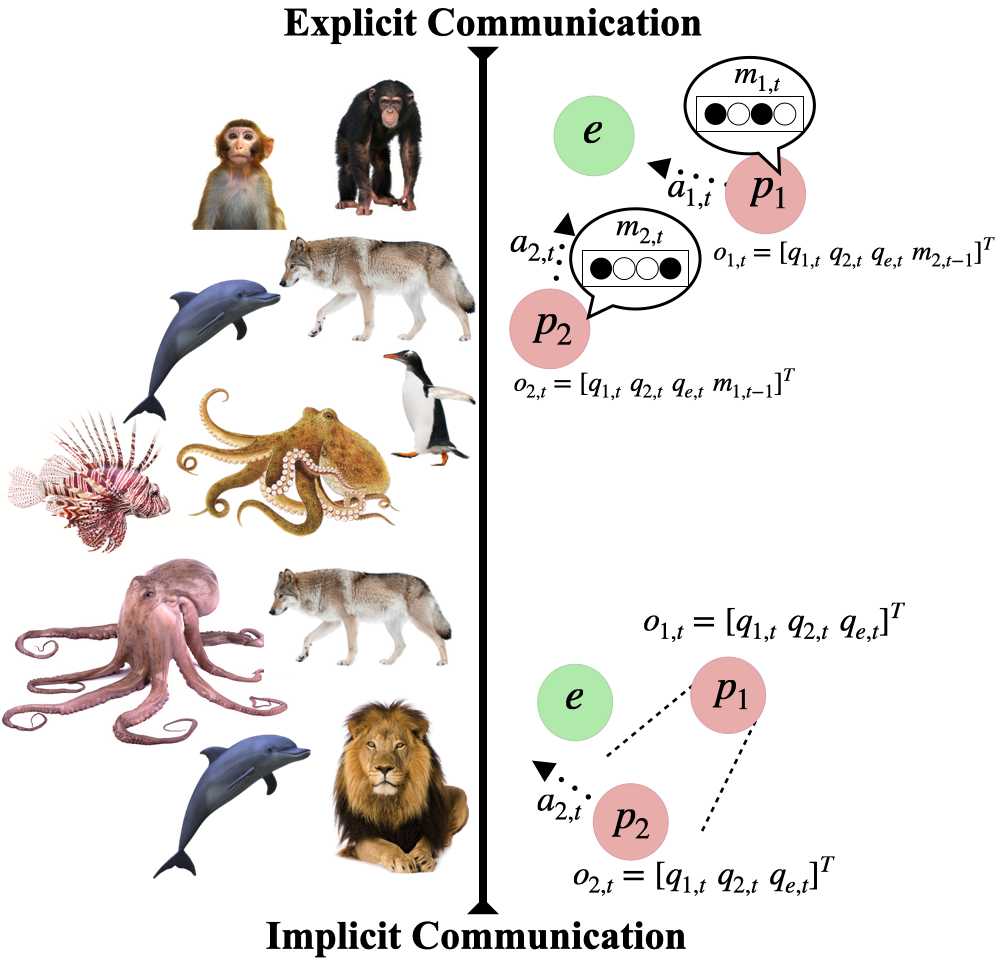}
  \caption{Left: Animal communication spans a complete spectrum from explicit to implicit communication. Right: A pursuit-evasion game with two pursuers $p_1$ and $p_2$ and an evader $e$ (with positions $q_{1,t}$, $q_{2,t}$, and $q_{e,t}$ at time $t$). Emergent communication only considers language-like communication (top), where agents share information over specialized sender-receiver architectures---e.g. $p_1$ receives the extended observation $o_{1,t} = [q_{1,t} \ q_{2,t} \ q_{e,t} \ m_{2, t-1}]^T$ and selects actions using the extra information $m_{2, t-1}$ that $p_2$ chose to share at the previous time-step (and vice versa for $p_2$). With implicit signals (bottom), $p_1$ selects a movement action $a_{1,t}$ based only on the observation $o_{1,t} = [q_{1,t} \ q_{2,t} \ q_{e,t}]^T$. The only information $p_1$ gets from $p_2$ is generated by $p_2$'s physical position (and vice versa for $p_2$). We posit that implicit signals are an important step towards learning richer communication.}
  \label{fig_spectrum}
\end{figure}

Though emergent communication is fundamentally an \textit{ab initio} approach compared to top-down language learning \cite{vaswani2017attention}, most recent methods at the intersection of emergent communication and multi-agent reinforcement learning have targeted protocols with complex structure and representational capacity, like that of human language \cite{lazaridou2020emergent, lowe2019pitfalls}. Such techniques equip agents with specialized sender-receiver architectures and allow them to exchange bits of information (or continuous signals) through these channels. One notable exception is the work of \citet{jaques2019social}, which considers how influential agents can pass information to other agents through their actions in the environment. We build on these ideas, recognizing the full spectrum of communication that exists in nature and proposing the study of lower-level forms of communication that do not require specialized architectures as a first step towards more sophisticated communication (see Figure \ref{fig_spectrum}).

Multi-agent cooperation in nature yields a wide range of communication protocols that vary in structure and complexity. In animal communication \cite{bradbury1998principles}, for example, reef-dwelling fish use body shakes \cite{vail2013referential, bshary2006interspecific} and octopuses punch collaborators \cite{sampaio2020octopuses}---forms of non-verbal communication---whereas chimps \cite{boesch1989hunting}, macaques \cite{mason1962communication}, and gentoo penguins \cite{choi2017group} each maintain a diverse vocal repertoire. Importantly, animals also exchange information through non-explicit channels. In the Serengeti, for example, lions use the posture of fellow pride members to stalk prey covertly \cite{schaller2009serengeti}. Moreover, wolves \cite{peterson2003wolf} and dolphins \cite{quick2012bottlenose}, both frequent vocal communicators, communicate implicitly during foraging---adjusting their group formation based on the position and orientation of other pack members \cite{herbert2016understanding}. \citet{breazeal2005effects} define such forms of communication, called \textit{implicit communication}, as "conveying information that is inherent in behavior but which is not deliberately communicated". Studies have shown that implicit cues are used frequently by humans as well. Human teams rely on gaze, facial expressions, and non-symbolic movement during cooperative tasks \cite{entin1999adaptive} including collaborative design \cite{wuertz2018design}, crowdsourcing \cite{zhou2017collaboration}, and multiplayer games (e.g. Hanabi \cite{liang2019implicit, bard2020hanabi}).

The role of implicit communication in teamwork has been studied extensively in the human-computer \cite{liang2019implicit, schmidt2000implicit} and human-robot interaction \cite{che2020efficient, breazeal2005effects, knepper2017implicit, butchibabu2016implicit} communities, and even in multi-agent systems \cite{pagello1999cooperative, gildert2018need}; but is less well-understood in the context of emergent agent-agent interactions \cite{de2010evolution}. This is due in large part to the anthropomorphized nature of standard definitions of implicit communication. Note that the key difference between explicit communication (a deliberate information exchange) and implicit communication (a non-deliberate exchange) in the definition above is \textit{intent} \cite{breazeal2005effects, gildert2018need}. Understanding implicit communication in the context of artificial agents requires (i) ascribing human-like mental states to artificial agents and (ii) giving agents the ability to reason about each other's mental states with respect to a common goal (\textit{\`a la} \citet{cohen1991teamwork}); which is a highly non-trivial undertaking.

In this work, we take a first step in this direction by examining the extent to which implicit communication aides coordination in multi-agent populations. Following the work of \citet{de2010evolution} in evolutionary settings, we study \textit{implicit signals}: signals generated by the physical position of an agent as it interacts with its environment. Implicit signals are observed passively by an agent's teammates, and so require neither special sender-receiver architectures nor reasoning about an agent's mental states. For RL agents, we distinguish implicit signals from explicit communication through each agent's actions and observations. As a simple example, consider the environment in Figure \ref{fig_spectrum} where, at time $t$, agents $p_1$, $p_2$, and $e$ are described by their positions $q_{1,t}$, $q_{2,t}$, and $q_{e,t}$ respectively. With implicit signals, $p_1$ selects a movement action $a_{1,t}$ based on the observation $o_{1,t} = [q_{1,t} \ q_{2,t} \ q_{e,t}]^T$---i.e. the only information $p_1$ gets from $p_2$ is the information contained in $p_2$'s physical position; and vice versa. With explicit communication, $p_1$ emits a communicative action $m_{1,t}$ in addition to $a_{1,t}$ (and same for $p_2$). In this case, $p_1$ receives the extended observation $o_{1,t} = [q_{1,t} \ q_{2,t} \ q_{e,t} \ m_{2, t-1}]^T$ and can select actions using the extra information $m_{2, t-1}$ that $p_2$ chose to share at the previous time-step.

Experimentally, we study coordination and the role of implicit communication with pursuit-evasion games \cite{isaacs1999differential}. Pursuit-evasion simulates an important coordination task, foraging, where communication is especially impactful. To accurately model the conditions under which coordination and communication emerges in the animal kingdom, we prioritize decentralized multi-agent learning. Despite the challenges posed by decentralized approaches \cite{foerster2016learning}, this feature is critical in the emergence of animal communications and so we focus on this important class of multi-agent learning frameworks to exploit inspiration from the natural world. Moreover, to encourage teamwork, we target pursuit-evasion games in which the pursuers are slower than the prey (coordination is not needed with the pursuers have a velocity advantage). Similar to \citet{lowe2017multi}, we study pursuit-evasion as a MARL problem. 

We show that, under these conditions, naive decentralized learning fails to solve the pursuit-evasion task. There are a number of challenges that lead to this outcome. First, decentralized learning is non-stationary \cite{lowe2017multi}---with multiple agents learning simultaneously, the learning problem is less stable for each agent individually. Next, since we prioritize tasks that require teamwork (e.g. pursuer speed less than evader speed), learning from scratch is difficult, as agents are forced to learn how to interact with the environment individually while simultaneously learning to coordinate with their teammates. Finally, pursuit-evasion is a sparse reward task---pursuers receive reward only when capturing the evader. During the early stages of training, when each agent is following a randomly initialized policy, it is virtually impossible for the agents to coordinate effectively enough to experience positive reward.

To address these challenges, we introduce a curriculum-driven approach for solving difficult, sparse reward coordination tasks. Our strategy, combining ideas from the ecological reinforcement learning \cite{co2020ecological} and automatic curriculum learning \cite{portelas2020automatic} literature, is motivated by the following insights: (i) we can create a \textit{curriculum over task difficulty by shaping the environment} to the current skill level of the cooperative agents; (ii) we can aide agents in learning multi-agent team behaviors by forcing them to \textit{learn single-agent behaviors first}. More specifically, we first adopt an environment shaping curriculum using velocity bounds, which allows pursuers to gather initial experience at a velocity greater than that of the evader, and then tune their strategies over time as velocity decreases. Then, we introduce a curriculum over agent behavior that warm-starts cooperative multi-agent learning by first seeding each agent with experience of successful single-agent behaviors.

Our first result confirms the importance of curriculum-driven multi-agent learning. We compare the performance of our strategy to ablations over the multi-agent curricula individually and show that, using our strategy, decentralized agents learn to solve difficult coordination tasks, such as the pursuit-evasion game.

Next, we stress-test both the strength of coordination and capacity for implicit communication of policies learned with our method. We compare the performance of multi-agent coordination learned with our method against a set of analytical and learned strategies that represent ablations over both coordination and implicit communication---i.e. outperforming one of these methods is equivalent to surpassing that method's level of sophistication in coordination, capacity for implicit communication, or both. 

Empirical results show that our method significantly outperforms highly sophisticated multi-agent strategies. Coordination strategies learned with our method successfully complete the pursuit-evasion task at a \textbf{speed ratio of 0.5} (i.e. the pursuers are moving at half the evader's speed), whereas each of the baseline methods fails to complete the pursuit-evasion task when pursuer speed is only slightly below evader speed (\textbf{speed ratio of 0.8)}.

Further, we hone in on the nature of each strategy's coordination more directly by examining agent behavior during the most important subsets of the trajectory---those that immediately precede the accumulation of reward. In the context of pursuit-evasion, we examine the distribution of pursuer locations relative the evader at the time of capture. By comparing rotational symmetry and rotational invariance in the agents' capture distributions, we find that our strategy learns structured coordination while simultaneously allowing pursuers to make dynamic adjustments to their position relative the evader to successfully achieve capture.

Finally, we examine the role of implicit signals (as introduced by \citet{de2010evolution}) more directly using measures of social influence. Influence measures such as Instantaneous Coordination \cite{jaques2019social} quantify the extent to which one agent's behavior influences its teammates. We repurpose this method to measure the exchange of implicit signals as \textit{position-based social influence}. Results of this study show that pursuers trained with our strategy exchange \textbf{up to 0.375 bits of information per time-step} on average, compared to a maximum of \textbf{only 0.15 bits on average} from the baseline methods. This indicates that our method learns, and relies heavily on, the exchange of implicit signals. 
\newline

\noindent \textbf{Preview of contributions}
Altogether, our work is summarized by the following contributions:
\begin{itemize}
    \item We highlight the importance of studying implicit signals as a form of emergent communication.
    \item We introduce a curriculum-driven learning strategy that enables cooperative agents to solve difficult coordination tasks with sparse reward.
    \item We show that, using our strategy, pursuers learn to coordinate to capture a superior evader, significantly outperforming sophisticated analytical and learned methods. Coordination strategies learned with our method complete the pursuit-evasion task at a speed ratio of $0.5$ (i.e. half the evader's speed), whereas each of the competing methods fails to complete the pursuit-evasion task at any speed ratio below $0.8$.
    \item We examine the use of implicit signals in emergent multi-agent coordination through position-based social influence. We find that pursuers trained with our strategy exchange up to 0.375 bits of information per time-step on average, compared to a maximum of only 0.15 bits on average from competing methods, indicating that our method has learned, and relies heavily on, the exchange of implicit signals.
\end{itemize}

\section{Related work}
\subsection{Emergent Communication}
Emergent communication examines the process by which cooperative agents learn communication protocols as a means to completing difficult multi-agent tasks \cite{lazaridou2020emergent}. Recent work has shown that MARL agents converge upon useful communication in referential games \cite{havrylov2017emergence} and can even develop language-like properties such as compositionality \cite{chaabouni2020compositionality,resnick2019capacity} and Zipf's Law \cite{chaabouni2019anti} when exposed to additional structural learning biases. More recently, this class of techniques has expanded to include complex situated environments \cite{das2019tarmac}, high-dimensional observations \cite{cowen2020emergent}, and the negotiation of belief states \cite{foerster2018bayesian}. Further work has shown that influential communication can be incentivized through additional learning objectives (i.e. inductive biases) \cite{eccles2019biases} and reward shaping \cite{jaques2019social}.

Implicit interactions are less well-studied in the emergent communication literature, despite their importance to teamwork, as outlined in the human-computer \cite{liang2019implicit, schmidt2000implicit} and human-robot interaction \cite{che2020efficient, breazeal2005effects, knepper2017implicit, butchibabu2016implicit} communities. Implicit communication has been studied for cooperative multi-agent tasks \cite{gildert2018need, grover2010implicit, pagello1999cooperative}, though not in the context of emergent behavior. Though some studies have shown that agents can learn to communicate non-verbally through actions and gestures \cite{mordatch2017emergence}, such forms of action-space communication \cite{baker2019emergent} are examples of non-verbal explicit communication, whereas our goal is to study completely implicit signals. Most similar to our work is that of \citet{de2010evolution}, which examines both implicit and explicit communication in evolutionary settings. Our work can be interpreted as bridging these ideas with the MARL literature.

\subsection{Multi-Agent Reinforcement Learning}
Multi-agent reinforcement learning (MARL) encompasses a large body of literature extending RL techniques to multi-agent settings. In general, MARL algorithms subscribe to either decentralized or centralized learning. With decentralized learning (or independent learning), each agent is responsible for updating its own policy (or Q-network) individually \cite{tan1993multi}. Though recent work has shown that decentralized learning is feasible in complex environments \cite{de2020independent}, it is often challenging, due to the non-stationary (and therefore unstable) nature of the learning problem \cite{laurent2011world, matignon2012independent}. Centralized techniques stabilize multi-agent learning by allowing agents to share a joint Q-network during training, then act independently at test-time \cite{lowe2017multi, foerster2018counterfactual}. Centralization has also proven useful for emergent communication, as agents can share gradients directly through their communication channels \cite{foerster2016learning}. Despite these benefits, we prioritize decentralized learning because it more accurately represents the learning problem that humans and animals face in the real-world.

Recently, significant effort has been aimed at connecting ideas from the curriculum learning literature \cite{bengio2009curriculum} to RL \cite{portelas2020automatic}. Such methods have derived curricula from virtually every aspect of the RL problem; including reward shaping \cite{bellemare2016unifying, pathak2017curiosity}, modifying initial state distributions \cite{florensa2017reverse}, and procedurally-generating sub-tasks \cite{portelas2020teacher, risi2019procedural}. Some multi-agent curricula have been shown to lead to more generally-capable RL agents \cite{team2021open}. Further work has examined the environment's role in generating curricula, leading to new methods such as environment shaping \cite{co2020ecological} and unsupervised environment design \cite{dennis2020emergent}. Of particular relevance is the work of \citet{co2020ecological}, which considers curricula that manipulate the dynamics of the RL environment to the benefit of learning agents. Our work combines a velocity-based variant of environment shaping with an additional curriculum for bootstrapping multi-agent learning with single-agent experience, which itself is inspired by the strategy employed by \citet{yang2018cm3} for multi-goal learning.

\subsection{Pursuit-Evasion}
Pursuit-evasion is a classic game setting for studying multi-agent coordination \cite{isaacs1999differential}. Though often played on a graph \cite{parsons1978pursuit}, work on continuous-space pursuit-evasion has enabled real-world applications such as unmanned aerial vehicles \cite{vidal2002probabilistic} and mobile robots \cite{chung2011search}. Further work has shown that optimal control strategies can be derived from value functions \cite{jang2005control} or even learned from scratch in MARL setting \cite{lowe2017multi}. A relevant class of pursuit-evasion games define the evader to be of equal or greater speed than the pursuers. This setting highlights the need for coordinated motion (e.g. encircling) \cite{vicsek2010closing} and communication \cite{wang2020cooperative} by the pursuers and is subject to theoretical performance bounds under these conditions \cite{ramana2017pursuit}. Our work uses this setting to study the emergence of animalistic coordination strategies and is similar to \citet{janosov2017group} in that regard, but with an emphasis on learned, rather than hand-defined, strategies.

\section{Preliminaries}
\label{sec_preliminaries}

\subsection{Markov Game}
A Markov game extends the Markov Decision Process (MDP) to multi-agent settings \cite{littman1994markov}. For $n$ agents, the formalism consists of a joint action-space $\boldsymbol{\mathcal{A}} = \{\mathcal{A}_1, ... , \mathcal{A}_n\}$ and joint observation space $\boldsymbol{\mathcal{O}} = \{\mathcal{O}_1, ... , \mathcal{O}_n\}$ that define the actions and observations, respectively. The environment defines a state space $\mathcal{S}$ from which a state $s_t$ is drawn each time-step. The environment transitions from its current state $s_t$ to a new state $s_{t+1}$ when actions are selected by each of the agents, as dictated by the transition function $T:\mathcal{S} \times \boldsymbol{\mathcal{A}} \rightarrow \mathcal{S}$. Finally, a reward function $r: \mathcal{S} \times \boldsymbol{\mathcal{A}} \rightarrow \mathbb{R}$ scores the quality of the composite action of the agents. The goal of each agent is to maximize the total reward it receives over time.

\subsection{Deep Deterministic Policy Gradients}
Deep Deterministic Policy Gradients (DDPG) is an off-policy actor-critic algorithm for policy gradient learning in continuous action spaces \cite{lillicrap2015continuous}. DDPG learns an optimal deterministic policy $\mu_\phi$ with respect to the RL objective:
\begin{equation}
    \label{eq_ddpg_obj}
    J(\phi) = \mathbb{E}_s[Q_{\omega}(s,a) |_{s=s_t, a=\mu_{\phi}(s_t)} ]
\end{equation}
\noindent by performing gradient ascent over the following gradient:
\begin{equation}
    \label{eq_ddpg_grad}
    \nabla_\phi J(\phi) = \mathbb{E}_s [\nabla_a Q_{\omega}(s,a)|_{s=s_t, a=\mu(s_t)}\nabla_\phi \mu(s)|_{s=s_t}]
\end{equation}
\noindent where $\phi$ and $\omega$ are parameters associated with $\mu$ and $Q$, respectively. Equation \eqref{eq_ddpg_grad} is a consequence of the deterministic policy gradient theorem \cite{silver2014deterministic}. For critic updates, DDPG minimizes the loss function:
\begin{equation}
    L(\omega) = \underset{s, a, r, s'}{\mathbb{E}} \big[\big(Q_\omega(s, a) - (r(s,a) + \gamma Q_\omega(s', \mu_\phi(s')))\big)^2\big]
\end{equation}
\noindent where ($s, a, r, s'$) are transition tuples sampled from a replay buffer and $\gamma {\in} [0,1]$ is a scalar discount factor. In this work, agents learn in a decentralized manner, each performing DDPG updates individually.

\subsection{Potential Field Navigation}
Given an agent $i$ with position $q_i$, we can define a potential function $U(q_i, q_{\textrm{goal}})$ between $q_i$ and a target point $q_{\textrm{goal}}$ such that the negative gradient $F(q_i, q_{\textrm{goal}}) = -\nabla U(q_i, q_{\textrm{goal}})$ specifies a control law for $i$'s motion. For example, let $U_{\textrm{att}}(q_i, q_{\textrm{goal}})$ be a quadratic function of distance between $q_i$ and a target point $q_{\textrm{goal}}$:
\begin{equation}
    \label{eq_attractive_potential}
    U_{\textrm{att}}(q_i, q_{\textrm{goal}}) = \frac{1}{2}k_{\textrm{att}} \, d(q_i, q_{\textrm{goal}})^2
\end{equation}
where $k_{\textrm{att}}$ is an attraction coefficient and $d(,)$ is a measure of distance. The resulting force exerted on agent $i$ is:
\begin{equation}
    F_{\textrm{att}} = -\nabla U_{\textrm{att}}(q_i, q_{\textrm{goal}}) = -k_{\textrm{att}}(q_i - q_{\textrm{goal}})
\end{equation}
\noindent In this work, the agent's action-space is defined in terms of headings, so only the \textit{direction} of this force impacts our agents.

\subsection{Pursuit-evasion}
Pursuit-evasion (i.e. predator-prey) is a classic setting for studying multi-agent coordination \cite{isaacs1999differential}. A pursuit-evasion game is defined between $n$ pursuers $\{p_1, ..., p_n \}$ and a single evader $e$. The goal of the pursuers is to catch the evader as quickly as possible and, conversely, the goal of the evader is to remain uncaught. Each agent $i$ is described by its current position and heading $q_i$ and is subject to planar motion $\dot{q}_i$:
\begin{align*}
    q_i =
    \begin{bmatrix}
        x_i\\
        y_i\\
        \theta_i
    \end{bmatrix}
    &&
    \dot{q}_i =
    \begin{bmatrix}
        \dot{x}_i\\
        \dot{y}_i\\
        \dot{\theta}_i
    \end{bmatrix}
    =
    \begin{bmatrix}
        \lvert \vec{v}_i \rvert \, \textrm{cos}(\theta_i) \\
        \lvert \vec{v}_i \rvert \, \textrm{sin}(\theta_i) \\
        \textrm{atan2}(\dot{y}_i, \dot{x}_i)
    \end{bmatrix}
\end{align*}
\noindent where $\vec{v}_i$ is the agent $i$'s velocity. The environment state $s_t$ is described by the position and heading of all agents $s_t = \{q_{p_1}, ..., q_{p_n}, q_e \}$. Upon observing $s_t$, each agent selects its next heading $\theta_i$ as an action. The chosen heading is pursued at the maximum allowed speed for each agent ($\lvert \vec{v}_p\rvert$ for the pursuers, $\lvert \vec{v}_e\rvert$ for the evader); with orientation changes being instantaneous. To encourage teamwork, we set $\lvert \vec{v}_p \rvert \leq \lvert \vec{v}_e\rvert$. We assume the evader to be part of the environment, as defined by the potential function:
\begin{equation}
    \label{eq_evader_objective}
    U_\textrm{evade}(\theta_e) = \sum_i \bigg(\frac{1}{r_i}\bigg) \cos(\theta_e - \tilde{\theta}_i)
\end{equation}
where $r_i$ and $\tilde{\theta}_i$ are the L2-distance and relative angle between the evader and the $i$-th pursuer, respectively, and $\theta_e$ is the heading of the evader. This objective is inspired by theoretical analysis of escape strategies in the pursuit-evasion literature \cite{ramana2017pursuit}. Intuitively, Equation \eqref{eq_evader_objective} pushes the evader away from pursuers, taking the largest bisector between any two when possible. The goal of the pursuers---to capture the evader as quickly as possible---is mirrored in the reward function, where $r(s_t, a_t) = 50.0$ if the evader is captured and $r(s_t, a_t) = -0.1$ otherwise. A derivation of Equation \eqref{eq_evader_objective} and additional environmental details are provided in Appendix \ref{sec_apdx_details}.

\subsection{Implicit Communication}
For clarity, we restate the definitions of implicit vs. explicit communication from \citet{breazeal2005effects} and implicit signals from \citet{de2010evolution}, which we adopt throughout this work.

\begin{definition}[Explicit Communication]
    \label{def_explicit_comm}
    Communication that is ``deliberate where the sender has the goal of sharing specific information with the collocutor" \cite{breazeal2005effects}.
\end{definition}
\begin{definition}[Implicit Communication]
    \label{def_implicit_comm}
    Communication that conveys ``information that [is] inherent in behavior but which is not deliberately communicated" \cite{breazeal2005effects}.
\end{definition}
\begin{definition}[Implicit Signal]
    \label{def_implicit_signal}
    The ``signal that is generated by the actual physical position of the [agents] and that is detected by the other [agents]" \cite{de2010evolution}.
\end{definition}

In the context of RL agents, an implicit signal refers to positional information observed by an agent's teammates as part of the environmental state $s_t$ or observation $o_t$ space, whereas explicit communication involves sending/receiving messages over a dedicated communication channel. We note that Definitions \ref{def_explicit_comm} and \ref{def_implicit_comm} rely on deliberation (or intention), whereas Definition \ref{def_implicit_signal} does not. For the purposes of our study---exploring the first-step in a bottom-up approach to emergent communication---we therefore focus on implicit signals as a form of emergent implicit communication.

\subsection{Instantaneous Coordination}
Instantaneous Coordination (IC) is a measure of social influence between agents \cite{jaques2019social}. IC is defined for two agents $i$ and $j$ as the mutual information $\textrm{I}(a_i^t; a_j^{t+1})$ between $i$'s action at time $t$ and $j$'s action at the next time-step. Formally, assuming agent $i$'s actions are drawn from the random variable $A_i$ with marginal distribution $P_{A_i}$ (and similarly for agent $j$), we can rewrite IC using the standard definition of mutual information as the Kullback-Leibler divergence between the joint distribution and the product of the marginals:
\begin{align*}
    I(A_i; A_j) &= D_{\textrm{KL}}(P_{A_i A_j} || P_{A_i} \times P_{A_j}) \\
    &= \sum_{\substack{a_i \in \mathcal{A}_i,\\ a_j \in \mathcal{A}_j}} P_{A_i A_j}(a_i, a_j) \log \bigg( \frac{P_{A_i A_j}(a_i, a_j)}{P_{A_i}(a_i) \times P_{A_j}(a_j)} \bigg)
\end{align*}
where $\mathcal{A}_i$ and $\mathcal{A}_j$ are the spaces over $A_i$ and $A_j$, respectively. Intuitively, high IC is indicative of influential behavior, while low IC indicates that agents are acting independently. We highlight that, in the absence of explicit communicative actions, IC is a measure of implicit signals (as defined in the previous section). This is because agent $i$'s action at time $t$ inherently dictates $i$'s position $q_i$ at time $t+1$, which is observed by $i$'s teammates. IC in this context is therefore a measure of position-based social influence. Since this work considers deterministic policies and uses IC at test-time, IC is computed over Monte-Carlo estimates of the relevant distributions.


\section{Method}
\label{sec_cd_ddpg}
Our goal is to learn multi-agent coordination with decentralized training and, in doing so, explore the role of implicit communication in teamwork. However, there are a number of issues with decentralized learning in difficult, sparse reward environments, such as our pursuit-evasion game. First, though setting $\lvert \Vec{v}_p \rvert \leq \lvert \Vec{v}_e\rvert$ is important for studying teamwork, it places the pursuers at a severe deficit. In the early stages of training---since each pursuer's action selection is determined by the randomly initialized weights of its policy network---the chance of slower pursuers capturing the evader defined in Equation \eqref{eq_evader_objective} is extremely low. The pursuers are unlikely to obtain a positive reward signal, which is vital to improving their policies. This issue is exacerbated by non-stationarity. In the case of decentralized DDPG, multiple agents learning in the same environment causes the value of state-action pairs for any one agent (as judged by its Q-function) to change as a result of policy updates of \textit{other} agents. This non-stationarity leads to higher-variance gradient estimates and unstable learning. Though recent advances in ``centralized training, decentralized execution" help in such cases \cite{lowe2017multi}, they violate our goal of decentralized learning.

To address these challenges, we introduce a \textit{curriculum-driven method} for decentralized multi-agent learning in sparse reward environments. Our strategy is motivated by the following principles: (i) we can create a \textit{curriculum over task difficulty by shaping the environment} to the current skill level of the cooperative agents; (ii) we can aide agents in learning multi-agent team behaviors by forcing them to \textit{learn single-agent behaviors first}. More specifically, we combine an environment shaping curriculum over agent velocities with a behavioral curriculum for bootstrapping cooperative multi-agent learning with successful single-agent experience. We refer to this curriculum-driven variant of DDPG as CD-DDPG throughout the rest of this paper.

\subsection{Velocity Ratio Curriculum}
Curriculum learning \cite{bengio2009curriculum} is a popular technique for solving complex learning problems by breaking them down into smaller, easier to accomplish tasks. Recent advances in ecological reinforcement learning \cite{co2020ecological} have introduced environment shaping---in which properties, initial states, or the dynamics of an environment are modified gradually---as a useful and more natural instantiation of curriculum learning for RL agents (than, say, traditional reward shaping). With this in mind, we construct a sequence of increasingly difficult pursuit-evasion environments by incrementally lowering the ratio of pursuer speed to evader speed ($\lvert \Vec{v}_p \rvert / \lvert \Vec{v}_e \rvert$). 

More formally, we define a curriculum over velocity bounds. Let $\Vec{v}_0$ be an initial setting of the environment's velocity ratio $\lvert \Vec{v}_p \rvert / \lvert \Vec{v}_e \rvert$. We anneal $\lvert \Vec{v}_p \rvert / \lvert \Vec{v}_e \rvert$ to a target ratio $\Vec{v}_{\textrm{target}}$ over $v_{\textrm{decay}}$ epochs as:
\begin{equation}
    \label{eq_curriculum}
    \Vec{v}_{i} \gets \Vec{v}_{\textrm{target}} + (\Vec{v}_0 - \Vec{v}_{\textrm{target}})*\max\bigg(\frac{(v_{\textrm{decay}}-i)}{v_{\textrm{decay}}}, 0.0\bigg)
\end{equation}
\noindent where $i$ represents the current training epoch and, in turn, $\Vec{v}_{i}$ the current velocity ratio. In practice, we initialize the environment such that $\Vec{v}_0 > 1.0$ (i.e. pursuers are faster than the evader), then anneal this ratio slowly as training progresses. This gives the pursuers an opportunity enjoy a velocity advantage early on in training, then develop increasingly coordinated strategies to capture the evader as $\lvert \Vec{v}_p \rvert / \lvert \Vec{v}_e \rvert$ decays.

\subsection{Behavioral Curriculum}
We combine the aforementioned velocity curriculum with a behavioral curriculum that extends off-policy learning to allow for targeted single-agent exploration early in the training process. As a consequence of the deterministic policy gradient and importance sampling \cite{silver2014deterministic}, the policy gradient can be estimated in an off-policy manner---i.e. using trajectories sampled from a separate behavior policy $\beta(a|s)$ where $\beta(a|s) \neq \mu_\phi$. Formally, this means Equation \eqref{eq_ddpg_grad} can be represented equivalently as:
\begin{align*}
    J_\beta(\phi) &= \int_\mathcal{S} p^\beta Q_{\omega}(s,a) |_{s=s_t, a=\mu_{\phi}(s_t)} \textrm{d}s
    \\[4pt]
    &= \mathbb{E}_{s \sim p^\beta}[Q_{\omega}(s,a) |_{s=s_t, a=\mu_{\phi}(s_t)} ]
\end{align*}
\noindent with the corresponding gradient:
\begin{equation*}
    \nabla_\phi J_\beta(\phi) = \mathbb{E}_{s \sim p^\beta} [\nabla_a Q_{\omega}(s,a)|_{s=s_t, a=\mu(s_t)}\nabla_\phi \mu(s)|_{s=s_t}]
\end{equation*}
\noindent where $p^\beta$ is the state distribution of the behavior policy $\beta$. 

Our behavioral curriculum takes advantage of off-policy learning by splitting training into two exploration phases that use distinct behavior policies $\beta_0$ and $\beta_{\mu}$, respectively. 
The key is that we define $\beta_0$ strategically to be a supervisory policy that collects \textit{successful single-agent experience} and $\beta_{\mu}$ to be a standard exploration policy ($\epsilon-$greedy for discrete actions, random noise for continuous actions). In multi-agent settings, defining $\beta_0$ this way allows each agent to learn how to interact with the environment first, before learning to coordinate with teammates. Without this curriculum, agents are forced to learn both simultaneously.

In our pursuit-evasion environment, we define $\beta_0$ as follows:
\begin{equation}
    \label{eq_beta_0}
     \beta_0 = -\nabla U_{\textrm{att}}(q_{p_i}, q_{e})
\end{equation}
\noindent where $U_{\textrm{att}}$ is the attractive potential-field defined in Equation \eqref{eq_attractive_potential}. Note that $\beta_0$ is a greedy policy that runs directly towards the evader. This strategy is obviously sub-optimal when $\lvert \Vec{v}_p \rvert / \lvert \Vec{v}_e \rvert \leq 1.0$, but helps pursuers learn to move in the direction of the evader when $\lvert \Vec{v}_p \rvert / \lvert \Vec{v}_e \rvert > 1.0$. In our experiments, we use $\beta_0$ only during the first phase of the velocity curriculum. After this phase, agents follow the standard DDPG behavior policy $\beta_\mu = \mu_\phi(s_t) + \mathcal{N}$, where $\mathcal{N}$ is the Ornstein-Uhlenbeck noise process. 

\section{Results}
\label{sec_results}
Our evaluation addresses two primary questions: (i) Does our curriculum-driven strategy enable decentralized agents to learn to coordinate in difficult sparse reward environments? (ii) To what extent does implicit communication emerge in the learned strategy? To answer these questions, we first perform an ablation study over each of the multi-agent curricula that comprise CD-DDPG. Then, we measure the performance of CD-DDPG against a set of analytical and learned strategies of increasing sophistication. We intentionally select these strategies to represent ablations over both coordination and implicit communication.

\subsection{Multi-Agent Curricula}
\label{sec_ablation}
\begin{figure}[]
    \centering
    \makebox[\linewidth][c]{\includegraphics[width=0.95\linewidth]{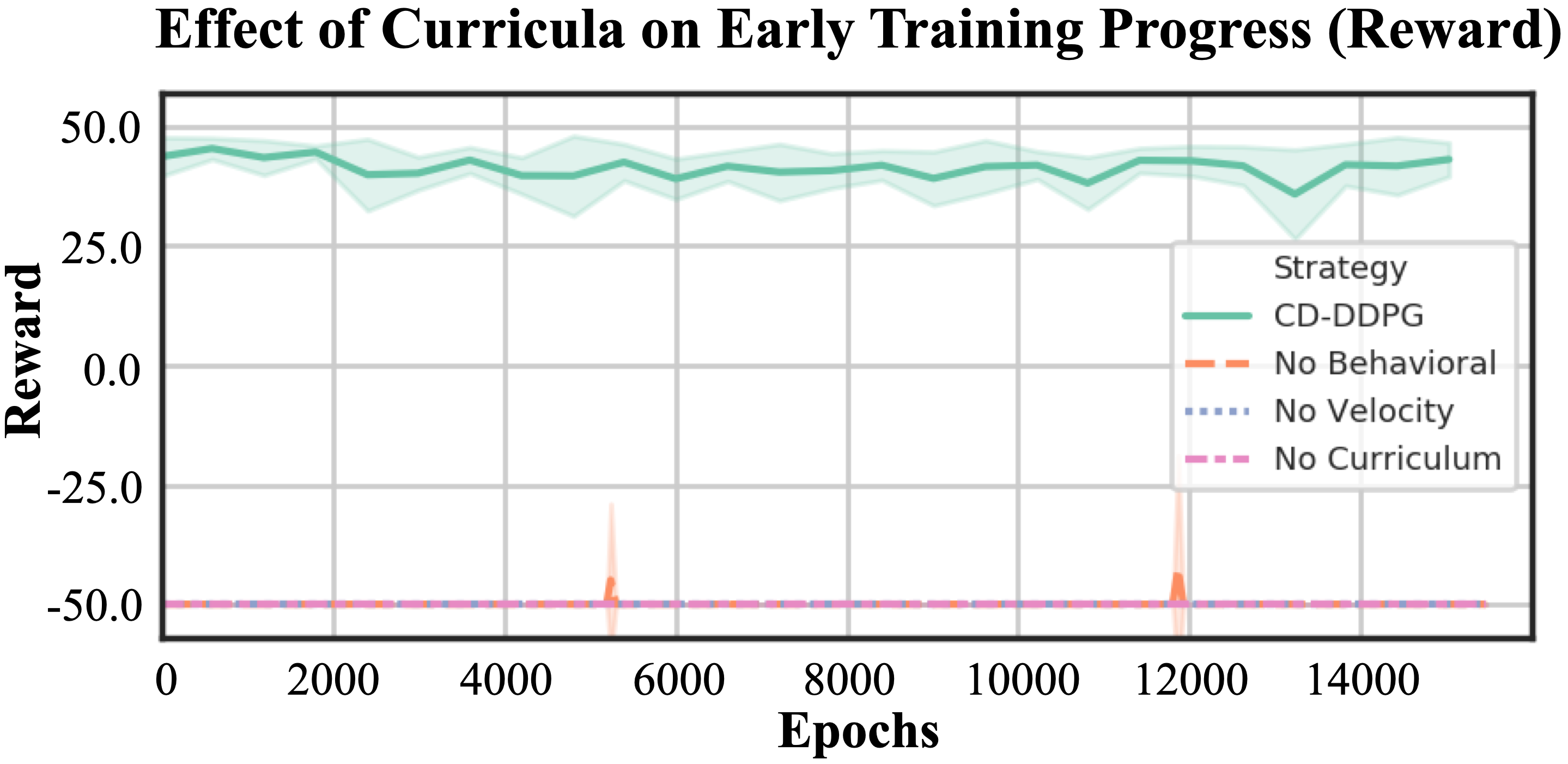}}
    \caption{An ablation study of curriculum-driven learning early in the training process. Note that $\lvert \Vec{v}_p \rvert / \lvert \Vec{v}_e \rvert$ is decreasing over time (increasing the task difficultly), so learning is characterized by \textit{sustained} reward rather than \textit{improved reward}. Only CD-DDPG successfully learns team coordination.}
    \label{fig_ablation}
\end{figure}
To isolate our method's performance, we run ablations over each of CD-DDPG's constituent parts. We compare CD-DDPG to the following alternatives: \textit{No Behavioral}, which follows the velocity curriculum, but not the behavioral one; \textit{No Velocity}, which follows the behavioral curriculum, but not the velocity one; and \textit{No Curriculum}, which uses neither curriculum during training (i.e. it is vanilla DDPG trained at a constant ratio $\lvert \Vec{v}_p \rvert / \lvert \Vec{v}_e \rvert = 0.7$). We trained each method for $15,000$ epochs in our pursuit-evasion environment across 10 different random seeds. Results are shown in Figure \ref{fig_ablation}.

\textit{No Curriculum}, unsurprisingly, flat-lines throughout the training process. It never experiences a positive reward signal, reflecting the difficulty of pursuit-evasion when $\lvert \Vec{v}_p \rvert / \lvert \Vec{v}_e \rvert < 1.0$. \textit{No Behavioral}, trained from a velocity ratio of $\lvert \Vec{v}_p \rvert / \lvert \Vec{v}_e \rvert = 1.2$ downwards, experiences small spikes in reward, but not consistently enough to improve the pursuers' policies. Finally, \textit{No Velocity} also fails to gain any traction (it's curve is behind No Curriculum in the figure), highlighting the importance of both curricula working together. This validates our intuition from Section \ref{sec_cd_ddpg}. Even with a velocity advantage, the agents struggle to learn individual interaction with the environment and coordination with their teammates simultaneously. \textit{CD-DDPG} is able to capture valuable experience even in the earliest stages of training. This bootstraps each pursuer's learning process, allowing them to maintain a high level of performance even after the warm-up is over. We therefore find that the combination of both curricula is crucial to warm-starting policy learning.

\subsection{Properties of Emergent Teamwork}
\begin{table}
  \caption{A summary of the strategies used to evaluate CD-DDPG and their capacity for coordination and implicit communication. Outperforming one of these methods is equivalent to surpassing that method's level of sophistication in coordination, capacity for implicit communication, or both.}
  \label{table_eval_straties}
  \centering
  \begin{tabular}{lll}
    \toprule
    Name     & Coordination     & Implicit Communication \\
    \midrule
    Greedy & \hspace{20pt}No & \hspace{40pt}No \\
    CD-DDPG (Partial) & \hspace{20pt}Yes & \hspace{40pt}No \\
    Pincer & \hspace{20pt}Yes & \hspace{40pt}Yes \\
    \bottomrule
  \end{tabular}
\end{table}
To study the strength of CD-DDPG's coordination further and identify the role of implicit communication in the agents' emergent teamwork, we compare the performance of CD-DDPG to a set of analytical and learned strategies of increasing sophistication. These strategies represent ablations over both coordination and implicit communication---i.e. outperforming one of these methods is equivalent to surpassing that method's level of sophistication in coordination, capacity for implicit communication, or both. With this in mind, we evaluate CD-DDPG against the following policies, which are also summarized in Table \ref{table_eval_straties}\footnote{We do not evaluate against centralized methods such as MADDPG \cite{lowe2017multi} because they violate our requirement of decentralized learning.}:

\paragraph{Greedy}
Each pursuer follows the greedy control strategy in Equation \eqref{eq_attractive_potential}. In Greedy pursuit, each pursuer ignores the positions of its teammates. Greedy pursuit therefore represents independent action (i.e. no coordination, no communication).

\paragraph{CD-DDPG (Partial)}
We train CD-DDPG under partial observability. Instead of the complete environment state $s_t$, each pursuer $p_i$ receives a private observation $o_t = \{q_{p_i}, q_e \}$ consisting of its own location $q_{p_i}$ and the location of the evader $q_e$. Despite not observing each other, CD-DDPG (Partial) pursuers are capable of coordinating through static role assignment. This is equivalent to assigning roles before each trajectory---i.e. $p_1$ always approaches from the left, $p_2$ from the right, etc.---and coordinating through these roles during pursuit. CD-DDPG (Partial) pursuers are therefore coordinated, but with no ability to communicate implicitly to modify their behavior extemporaneously.

\paragraph{Pincer}
We define the Pincer strategy as an adversarial function that exploits knowledge of the evader's objective in Equation \eqref{eq_evader_objective}:
\begin{align}
    \label{eq_pincer}
    F(\boldsymbol{\tilde{\theta}_i, r_i}) &= \underset{\boldsymbol{\tilde{\theta}_i, r_i}}{\max} \big[ \underset{\theta_e}{\min} \big[ U_\textrm{evade}(\theta_e)\big] \big] \\
    &= \underset{\boldsymbol{\tilde{\theta}_i, r_i}}{\max} \bigg[ \underset{\theta_e}{\min} \bigg[ \sum_i \bigg(\frac{1}{r_i}\bigg) \cos(\theta_e - \tilde{\theta}_i) \bigg] \bigg]
\end{align}
\noindent where $\boldsymbol{\tilde{\theta}_i}$ and $\boldsymbol{r_i}$ are the polar coordinates of each pursuer relative the evader. Intuitively, Pincer encircles the evader and cuts off potential bisector escape paths while enclosing the circle. It therefore supports both coordination---pursuers uphold a circular formation---and implicit communication---every time-step, pursuers use information from the locations of their teammates to adjust their own position on the circular formation. We provide additional details on the Pincer strategy in Appendix \ref{sec_apdx_details_pincer}.

All experiments involve $n=3$ pursers. For CD-DDPG and CD-DDPG (Partial), agents are trained for $50,000$ epochs. The velocity ratio is decreased from $\lvert \Vec{v}_p \rvert / \lvert \Vec{v}_e \rvert = 1.2$ to $\lvert \Vec{v}_p \rvert / \lvert \Vec{v}_e \rvert = 0.4$ in decrements of $0.1$ (i.e. eight separate training sessions each). Velocity ratio decay occurs over $v_{\textrm{decay}}=15000$ epochs, during which the behavioral curriculum $\beta_0$ is used (but only during the first training session). After training, we test the resulting policies at each velocity step. Test-time performance is measured over 100 independent trajectories (averaged over five random seeds each).

\subsubsection{Capture Success}
\begin{figure}
  \centering
  \includegraphics[width=.99\linewidth]{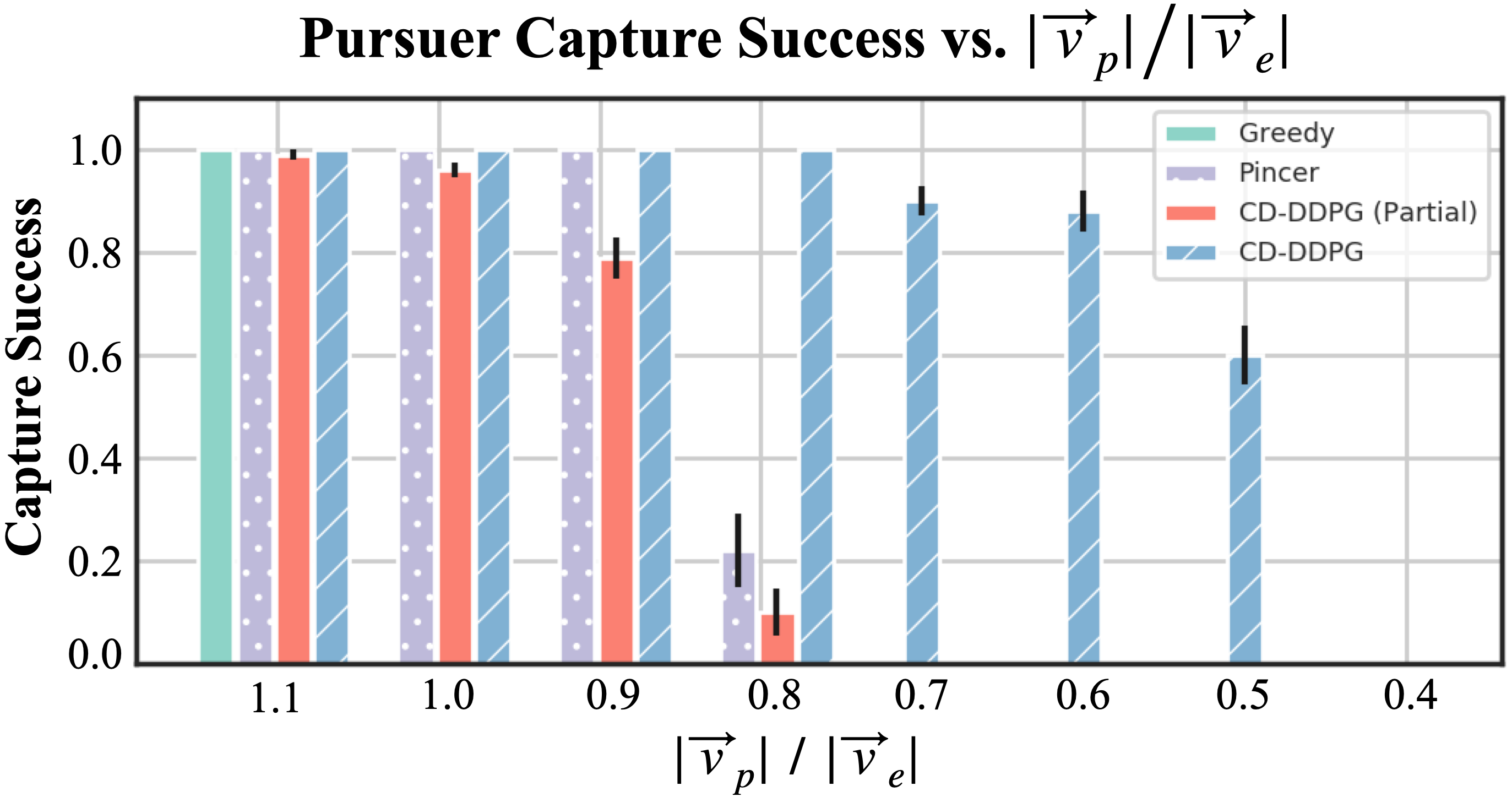}
  \caption{Capture success as a function of velocity. CD-DDPG succeeds at $\lvert \Vec{v}_p \rvert / \lvert \Vec{v}_e \rvert = 0.5$ whereas each competing method fails below $\lvert \Vec{v}_p \rvert / \lvert \Vec{v}_e \rvert = 0.8$.}
  \label{fig_cap_success}
\end{figure}
First, we evaluate capture success as a function of the velocity advantage of the evader. The results are shown in Figure \ref{fig_cap_success}. Unsurprisingly, each method has a high capture success rate when $\lvert \Vec{v}_p \rvert / \lvert \Vec{v}_e \rvert > 1.0$. The Greedy strategy drops off at $\lvert \Vec{v}_p \rvert / \lvert \Vec{v}_e \rvert = 1.0$, which is also expected---a straight-line chase only works when $\lvert \Vec{v}_p \rvert / \lvert \Vec{v}_e \rvert > 1.0$. The Pincer and CD-DDPG (Partial) strategies are able to coordinate successfully at lower speeds, but eventually \textbf{fail to capture the evader at $\boldsymbol{\lvert \Vec{v}_p \rvert / \lvert \Vec{v}_e \rvert = 0.8}$} and below. As $\lvert \Vec{v}_p \rvert / \lvert \Vec{v}_e \rvert$ decreases further, CD-DDPG significantly outperforms the other strategies. Specifically, CD-DDPG successfully completes the pursuit-evasion task at a \textbf{speed ratio of 0.5} (i.e. pursuers moving at half the evader's speed). These results show that CD-DDPG learns to coordinate significantly more effectively than the other strategies.

\subsubsection{Relative position during capture}
\begin{figure}[t!]
    \centering
    \makebox[\linewidth][c]{\includegraphics[width=0.85\linewidth]{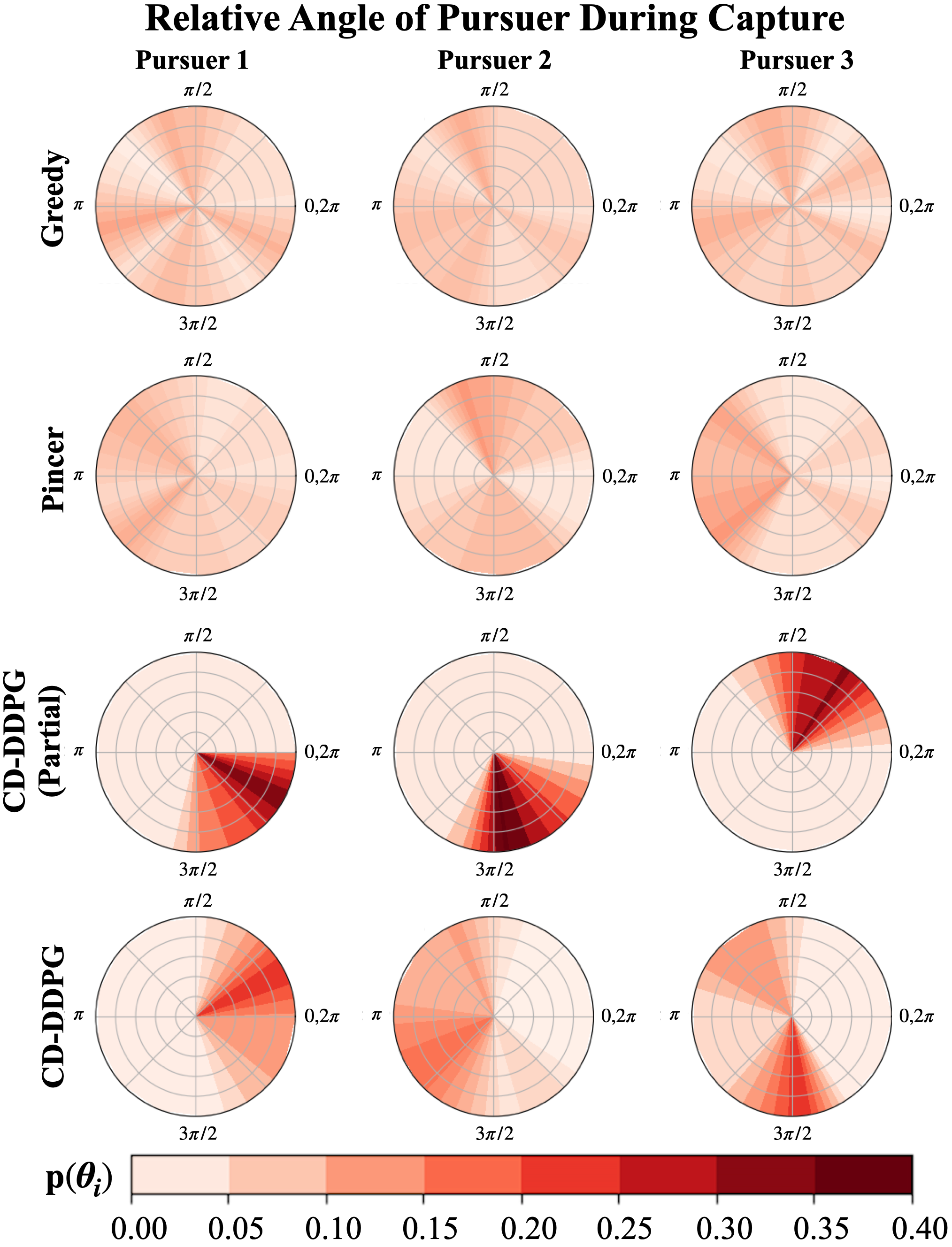}}
    \caption{Distribution of relative angle for each pursuer during capture, binned in the range $[0, 2\pi]$ and displayed as a heatmap. Each row represents a pursuit strategy.}
    \label{fig_capture_pos}
\end{figure}
Next, we study the nature of each strategy's coordination with finer granularity. In sparse reward tasks, the most important time-steps are those that immediately precede reward. For this reason, we look at the distribution of pursuer locations at the time of capture. We collect 100 successful trajectories from each strategy and compute the distribution of pursuer positions relative the evader. In general, we expect coordinated pursuit to exhibit rotational symmetry during capture. Rotational symmetry suggests that pursuers have learned strategies which lead them to well-defined ``capture points" around the evader. Conversely, rotational invariance is indicative of independent pursuit---i.e. pursuers do not follow concrete patterns of attack. Results for this study are shown in Figure \ref{fig_capture_pos}.

We find that Greedy and Pincer both yield uniform capture distributions. This is unsurprising for Greedy pursuers, whose pursuit paths are not effected by their teammates. The Pincer strategy encircles the evader, but does not constrain pursuers to specific locations on the circle. This leads to less-structured capture(i.e. weaker role assignment) by the Pincer pursuers. In contrast, CD-DDPG (Partial) pursuers demonstrate very strong role assignment, with each pursuer capturing the evader from the same relative angle each time. Taken into context with the results from Figure \ref{fig_cap_success}, it is clear that this level of rotational symmetry impacts success. In fact, it is an example of over-commitment to role assignment. The pursuers adopt very constrained roles---e.g. ``$p_1$ always move left", ``$p_2$ always move right"---which works when $\lvert \Vec{v}_p \rvert / \lvert \Vec{v}_e \rvert \geq 1.0 $, but fails at lower velocities. CD-DDPG balances rotational symmetry and invariance. Each pursuer follows a unique angle towards the evader, but does not commit to that angle completely. CD-DDPG therefore learns structured coordination while allowing pursuers to make dynamic adjustments to their position relative the evader to achieve capture.

\subsubsection{Position-based Social influence}
\begin{figure}
  \centering
  \includegraphics[width=.9\linewidth]{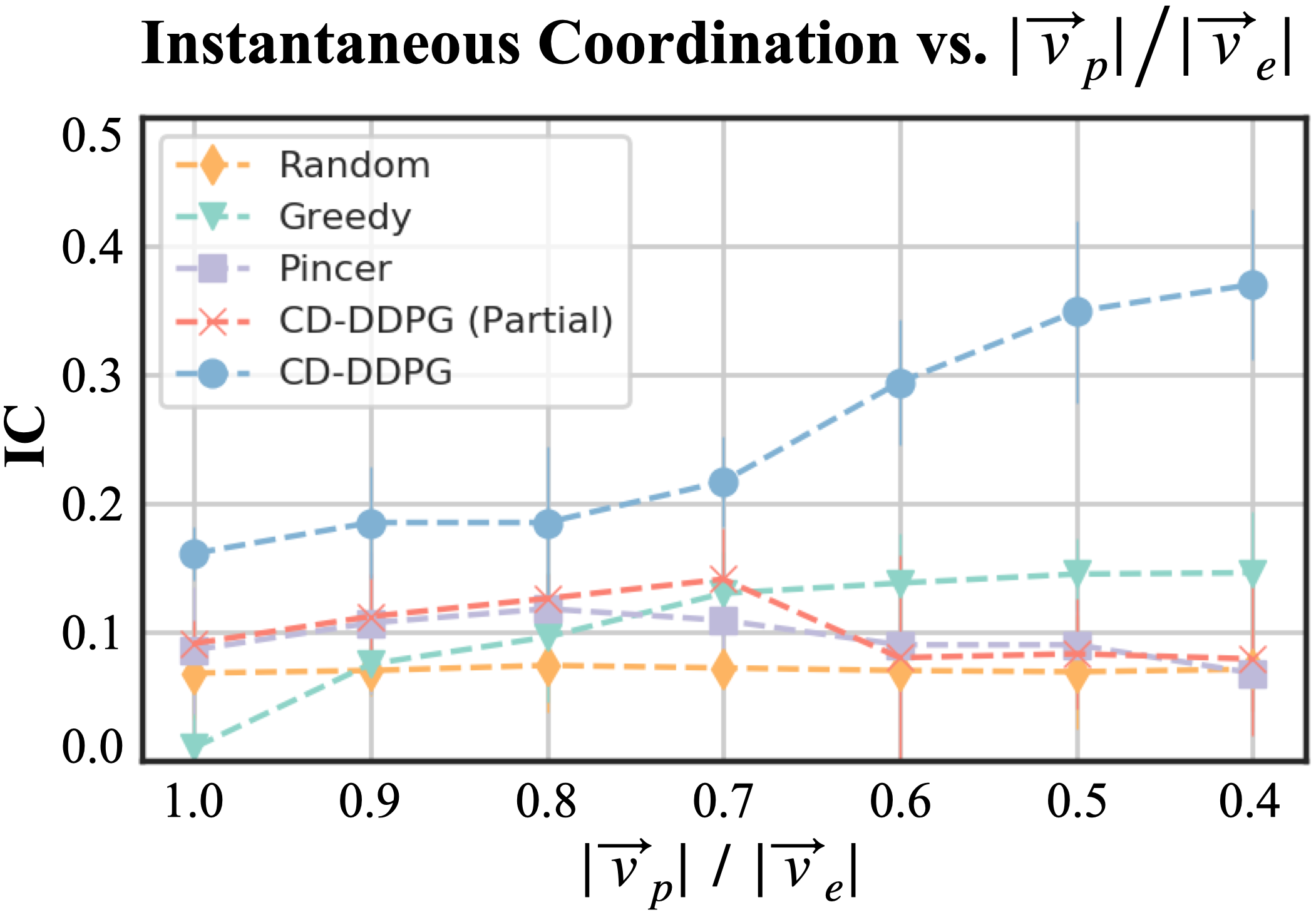}
  \caption{Instantaneous Coordination (IC) results as a function of velocity ratios. Agents trained with CD-DDPG exchange up to 0.375 bits of information per time-step on average, increasing as $\lvert \Vec{v}_p \rvert / \lvert \Vec{v}_e \rvert$ drops, whereas competing strategies peaks remains constant.}
  \label{fig_ic_results}
\end{figure}
To further study the role of implicit communication in pursuer performance, we compute the IC score for each strategy. As noted in, section \ref{sec_preliminaries}, by measuring the amount that one agent's actions (and therefore its next position) influences the actions of its teammates, IC quantifies the exchange of implicit signals (see Definition \ref{def_implicit_signal}) amongst teammates. Following \citet{jaques2019social}, we compute IC empirically as a Monte-Carlo approximation over multi-agent trajectories. We average influence across all trajectory steps and for each agent-agent pair. We also evaluate pursuers that act randomly, which provides a baseline for independent action. The results are shown in Figure \ref{fig_ic_results}.

We find that, as $\lvert \Vec{v}_p \rvert / \lvert \Vec{v}_e \rvert$ decreases, the IC levels attained by CD-DDPG increases significantly, whereas it remains stagnant for other methods. In fact, across all $\lvert \Vec{v}_p \rvert / \lvert \Vec{v}_e \rvert$ levels, we find that pursuers trained with \textbf{CD-DDPG exchange up to 0.375 bits of information} per time-step on average, compared to a maximum of \textbf{only 0.15 bits on average} from the baseline methods. This indicates that CD-DDPG \textit{achieves increasingly complex coordination} as task difficulty increases and is a promising sign that the CD-DDPG team is exchanging implicit signals---i.e. each pursuer is responding to positional information from its teammates. Finally, we note a minor (though surprising) increase in coordination for the greedy pursuers at low velocities. This is an artifact we dub ``phantom coordination" and discuss in the Appendix \ref{sec_apdx_qual_phantom}.

\subsubsection{High-influence moments}
\begin{figure}[t!]
    \centering
    \makebox[\linewidth][c]{\includegraphics[width=0.9\linewidth]{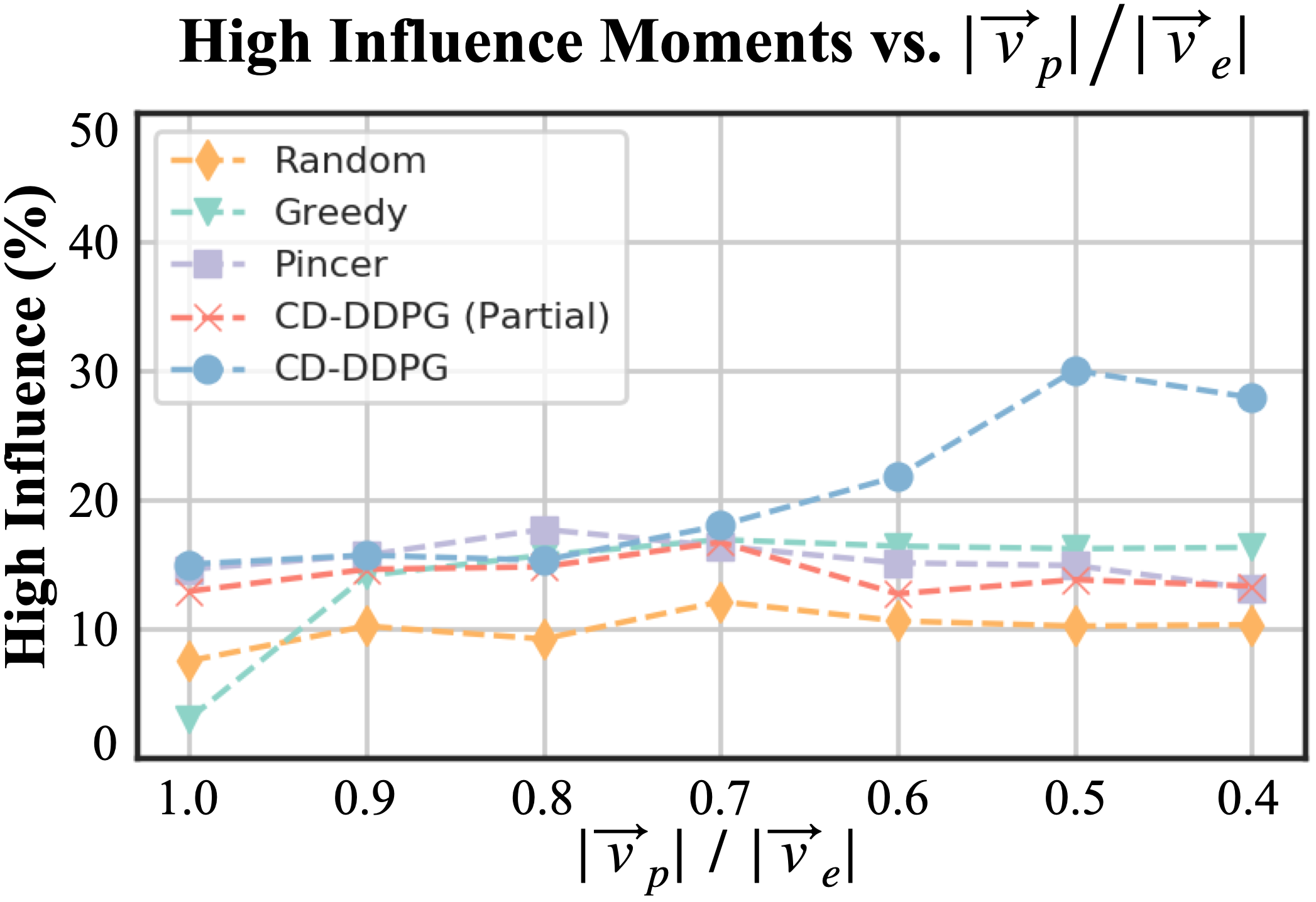}}
    \caption{Percentage of high-influence moments as a function of velocity ratios. Agents trained with CD-DDPG become more influential as $\lvert \Vec{v}_p \rvert / \lvert \Vec{v}_e \rvert$ decreases.}
    \label{fig_high_ic}
\end{figure}
We also report on the percentage of high-influence moments that occur between pairs of agents (see Figure \ref{fig_high_ic}). A high-influence moment is a time-step in which IC is above the mean IC for the entire trajectory. Similar to previous work \cite{jaques2019social}, we find that influence is sparse in most cases---\textbf{only between 10-15\%} of trajectory steps exhibit high-influence across all $\lvert \Vec{v}_p \rvert / \lvert \Vec{v}_e \rvert$ levels. The exception, notably, occurs for CD-DDPG. At low speeds (i.e. $\lvert \Vec{v}_p \rvert / \lvert \Vec{v}_e \rvert \leq 0.7$), we see a significant increase in the percentage of high-influence moments between the CD-DDPG pursuers, reaching a \textbf{maximum near 30\%}. This is further evidence that, as $\lvert \Vec{v}_p \rvert / \lvert \Vec{v}_e \rvert$ decreases, CD-DDPG pursuers form increasingly highly-coordinated formations and make split-second decisions based on the movements of their teammates. This points more concretely to the use of implicit signals between CD-DDPG pursuers.

Interestingly, the behavior of CD-DDPG closely matches the documented behaviors of social predators such as dolphins and wolves—i.e. sudden changes of position/orientation as a response to the movements of other teammates \cite{herbert2016understanding}. We elaborate on these and other qualitative findings in Appendix \ref{sec_apdx_qual}.
\section{Conclusion}
\label{sec_conclusion}
This work connects emergent communication to the spectrum of communication that exists in nature, highlighting the importance of interpreting communication as a spectrum from implicit to explicit communication. We proposed a curriculum-driven strategy for policy learning in difficult multi-agent environments. Experimentally, we showed that our curriculum-driven strategy enables pursuers to coordinate and capture a superior evader, outperforming other highly-sophisticated analytic pursuit strategies. We also provided evidence suggesting that the emergence of implicit communication is a key contributor to the success of this strategy. There are a number of extensions of this work that study how common principles contribute to integrated, communicative behavior; including: imperfect state information, increased environmental complexity, and nuanced social dynamics between agents.






\bibliographystyle{ACM-Reference-Format} 
\bibliography{sample}

\clearpage

\appendix

\section{Pursuit-Evasion Details}
\label{sec_apdx_details}

\subsection{Pursuit-Evasion Environment}
\begin{figure}[]
    \centering
    \makebox[\linewidth][c]{\includegraphics[width=0.99\linewidth]{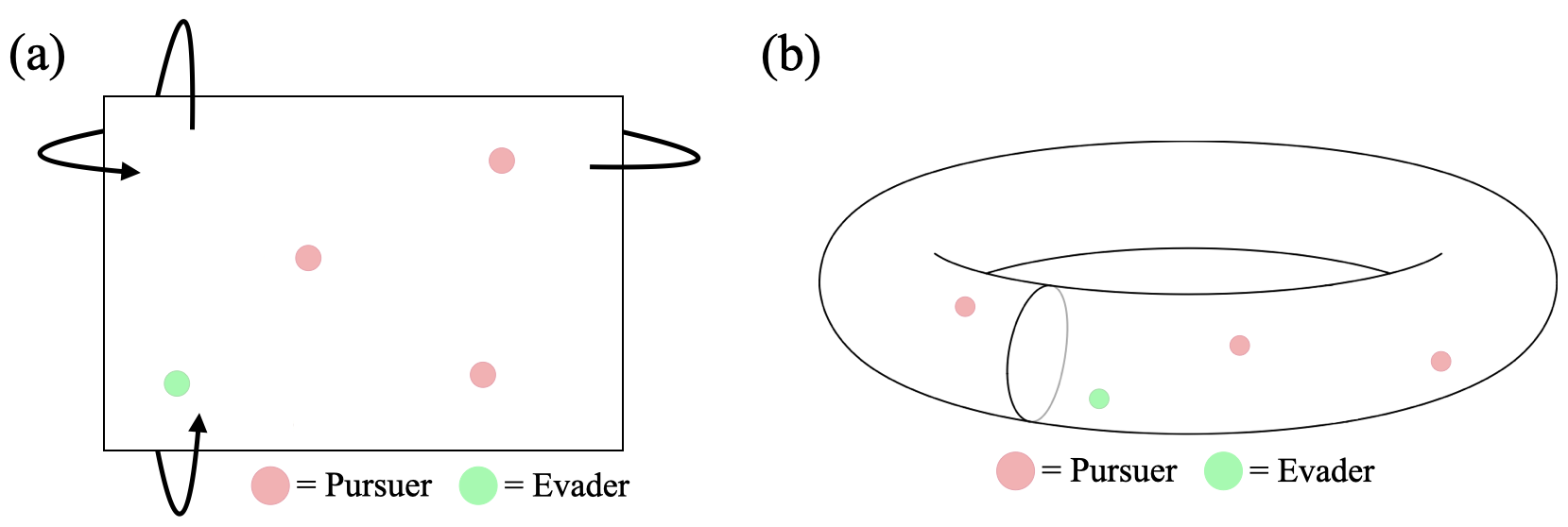}}
    \caption{A planar pursuit-evasion game with periodic boundary conditions interpreted as a toroidal pursuit-evasion environment.}
    \label{fig_torus_env}
\end{figure}
Our environment is a toroidal extension of the planar pursuit-evasion game proposed by \citet{lowe2017multi}. In general, unbounded planar pursuit-evasion can be described by two cases:
\begin{itemize}
    \item Case 1: $\lvert \Vec{v}_p \rvert > \lvert \Vec{v}_e\rvert$. The game is solved by a straight-line chase towards the evader and is not interesting from the perspective of coordination.
    \item Case 2: $\lvert \Vec{v}_p \rvert \leq \lvert \Vec{v}_e\rvert$. The evader has a significant advantage. Pursuers have at most one opportunity to capture the evader and are usually only successful under strict initialization conditions \cite{ramana2017pursuit}.
\end{itemize}
\noindent \citet{lowe2017multi} addressed this by penalizing agents for leaving the immediate area defined by the camera with negative reward. The evader defined by Equation \eqref{eq_evader_objective}, however, will run away indefinitely in the $\lvert \Vec{v}_p \rvert \leq \lvert \Vec{v}_e\rvert$ case. To provoke consistent interaction between agents, we extend the planar environment with periodic boundary conditions. One can think of this as playing the pursuit-evasion game on a torus (see Figure \ref{fig_torus_env}).

We argue that toroidal pursuit-evasion is better suited for our problem setting. First, it does not require strict initialization conditions or special rewards. We can initialize pursuers randomly and allow them to construct ad-hoc formations. Second, pursuit is no longer a one-and-done proposition. This reflects the notion that, in nature, predators often do not give up after a single attempt at a prey---they regroup and pursue it again.

\subsection{Evader Strategy Explained}
\subsubsection{Derivation}
The goal of the evader strategy in Equation \eqref{eq_evader_objective} is to run from pursuers along the maximum bisector between two pursuers. Given pursuer positions $\{q_{p_1}, ..., q_{p_n}\}$, we compute polar coordinates:
\begin{align*}
    r_i &= d(q_e, q_{p_i})
    \\
    \tilde{\theta}_i &= \textrm{atan2}(y_{p_i}, x_{p_i})
\end{align*}

\noindent for each pursuer $p_i$ relative the evader. Next, we define a potential field that will push the evader towards a bisector:
\begin{equation*}
    U(\theta_e) = \sum_i \cos(\theta_e - \tilde{\theta}_i)
\end{equation*}
\noindent Using Ptolemy's difference formula, we can expand the potential field as:
\begin{align*}
    U(\theta_e) &= \sum_i \cos(\theta_e - \tilde{\theta}_i) \\
    &= \sum_i \cos(\theta_e)\cos(\tilde{\theta}_i) + \sin(\theta_e)\sin(\tilde{\theta}_i) \\
    &= A \cos(\theta_e) + B \sin(\theta_e)
\end{align*}
\noindent when we plug-in the known $\tilde{\theta}_i$ values. The function $U(\theta_e)$ is maximized/minimized for values of $A$ and $B$ such that:
\begin{equation*}
    \nabla U(\theta_e) = -A\sin(\theta_e) + B\cos(\theta_e) = 0
\end{equation*}
\noindent which simplifies to:
\begin{equation*}
     \tan(\theta_e) = \frac{B}{A}
\end{equation*}
\noindent The evader follows the direction of the negative gradient ($-\nabla U(\theta_e)$) and pursues it at maximum speed. Modulating the cost function by $r_i$:
\begin{equation*}
    U(\theta_e) = \sum_i \bigg(\frac{1}{r_i}\bigg) \cos(\theta_e - \tilde{\theta}_i)
\end{equation*}
\noindent allows the evader to modify its bisector based on the distance to each pursuer. This helps significantly when the evader is stuck in symmetric formations.

\subsubsection{Unit Tests}
\begin{figure}[]
    \centering
    \makebox[\linewidth][c]{\includegraphics[width=0.95\linewidth]{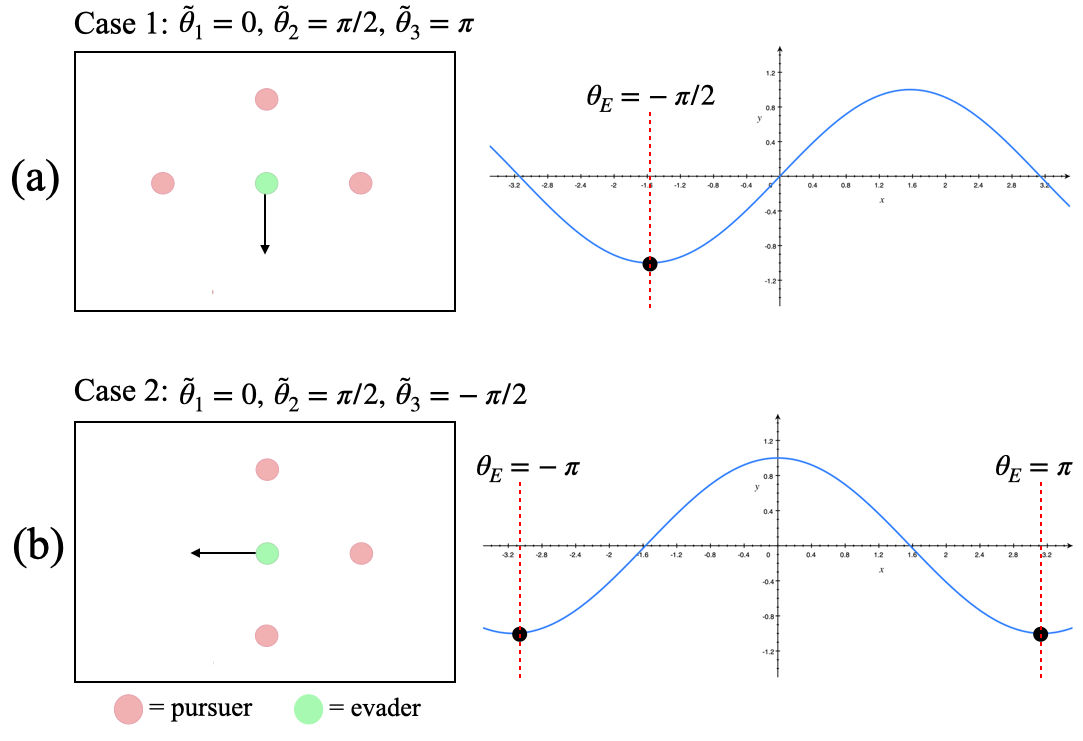}}
    \caption{Unit tests simulating one time-step of action selection as dictated by the evader's cost function. (a) Given pursuers at relative angles $\tilde{\theta}_1 = 0, \tilde{\theta}_2=\pi/2, \tilde{\theta}_3=\pi$ the evader will select the heading $\theta_e = -\pi/2$, which minimizes its cost. (b) Similarly, the evader will select the heading $\theta_e = -\pi = \pi$ when pursuers are located at relative angles $\tilde{\theta}_1 = 0, \tilde{\theta}_2=\pi/2, \tilde{\theta}_3=-\pi/2$.}
    \label{fig_unit_test}
\end{figure}
We include a set of ``unit tests" that shed light on the evader's decision-making behavior. We assume $n=3$ pursuers are stationed around the evader at relative angles $\tilde{\theta}_1$, $\tilde{\theta}_2$, and $\tilde{\theta}_3$. For simplicity, we initialize the pursuers such that $\forall{i}, r_i = 1$ to negate the effects of radius modulation.
\begin{itemize}
    \item \underline{Case 1:} $\tilde{\theta}_1 = 0, \tilde{\theta}_2=\pi/2, \tilde{\theta}_3=\pi$.
    Pursuers are spaced equally around the upper-half of the unit circle. In this case, the cost minimizer occurs for $\theta_e = -\pi/2$ (see Figure \ref{fig_unit_test}a).
    \item \underline{Case 2:} $\tilde{\theta}_1 = 0, \tilde{\theta}_2=\pi/2, \tilde{\theta}_3=-\pi/2$.
    Pursuers are spaced equally around the right-half of the unit circle. In this case, the cost minimizer occurs for $\theta_e = -\pi$ and $\theta_e = \pi$(see Figure \ref{fig_unit_test}b). Either solution can be selected to move the evader towards the largest opening.
\end{itemize}

\noindent In general, the cosine function imposes structure on the evader's objective---it will oscillate between $[-1, 1]$ over a period of $\pi$, taking on a maximum value of $U(\theta_e) = 1$ when the difference between the evader's heading $\theta_e$ and the relative angle of a pursuer $\tilde{\theta}_i$ is zero and a minimum $U(\theta_e) = -1$ when $\theta_e - \tilde{\theta}_i = \pi$. Summing over all $\tilde{\theta}_i$'s incentivizes the evader to follow the heading that splits the largest bisector of the pursuers, as shown in the examples.

\subsection{Pincer Pursuit Explained}
The Pincer strategy described by Equation \eqref{eq_pincer} is inspired by prior work on theoretical pursuit-evasion \cite{ramana2017pursuit}. Solving Equation \eqref{eq_pincer} requires optimizing over both $\boldsymbol{\tilde{\theta}_i}$ and $\boldsymbol{r_i}$. Fortunately, we can exploit the toroidal structure of the environment to construct an optimization routine that solves for $\boldsymbol{\tilde{\theta}_i}$ and $\boldsymbol{r_i}$ discretely. In particular, we can unroll the torus $k$ steps in each direction to generate $(2k + 1)^2$ replications of the current environment state. Rather than solving for optimal $\boldsymbol{\tilde{\theta}_i}$ and $\boldsymbol{r_i}$ values directly, we find the set $\boldsymbol{P}$ of pursuers that maximize Equation \eqref{eq_pincer} across all replications of the environment. We constrain the problem by limiting selections of each pursuer $p_i$ to replications of \textit{itself only}. This dramatically cuts down the number of possible sets $\boldsymbol{P}$ from $\binom{(2k+1)^2n}{n}$ to $\binom{(2k+1)^2}{1} \cdot \binom{(2k+1)^2}{1} \cdot \binom{(2k+1)^2}{1}$, where $n$ is the number of pursuers in the environment. Thus, we solve Equation \eqref{eq_pincer} via a discrete optimization over each of the $((2k + 1)^2)^3$ possible pursuer selections.
\label{sec_apdx_details_pincer}
\begin{figure}[]
    \centering
    \makebox[\linewidth][c]{\includegraphics[width=0.8\linewidth]{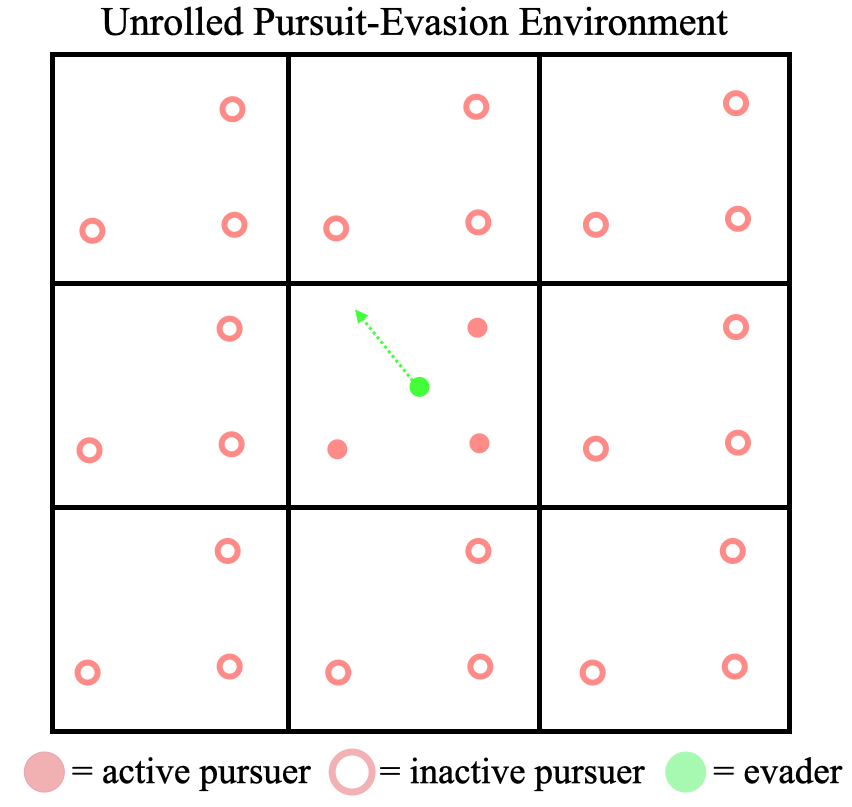}}
    \caption{The torus environment unrolled $k=1$ times in each direction. The filled in red circles denote the ``active" pursuers that are pursuing the evader at the current time-step, whereas the empty circles represent ``inactive" pursuers. We consider only a single evader, located in the center tile.}
    \label{fig_replica}
\end{figure}
The resulting set $\boldsymbol{P}$ defines the set of ``active" pursuers that will pursue the evader directly at the next time-step. Due to the nature of the evader's objective function---it is attracted to bisectors and repulsed from pursuers---the maximum $\boldsymbol{P}$ tends to favor symmetric triangular formations. Though this method obviously does not scale well with $n$ and $k$, we found that we are able to find a sufficient maximizer with low values of $k$ (i.e. $k=1$ in our experiments). The replication process is shown for the $k=1$ case in Figure \ref{fig_replica}. Note that we discriminate between ``active" pursuers---i.e. those $p_i \in P$ pursuing the evader at the current time-step---from ``inactive" pursuers.

\section{Qualitative Results}
\label{sec_apdx_qual}
\begin{figure}[t!]
    \centering
    \makebox[\linewidth][c]{\includegraphics[width=0.95\linewidth]{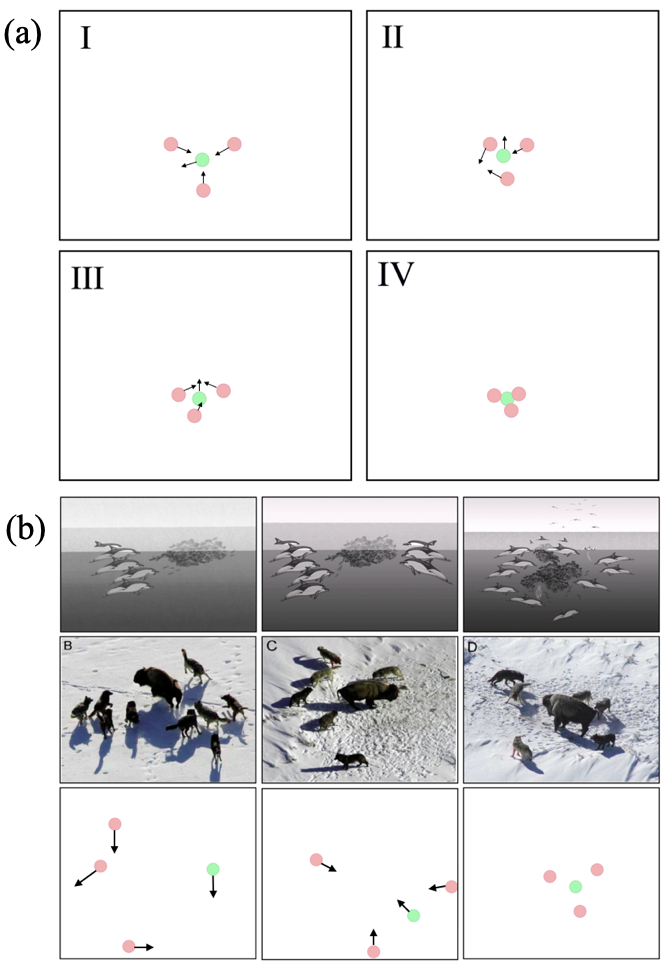}}
    \caption{Qualitative results from the pursuit-evasion experiment. (a) The pursuers coordinate to capture the evader, displaying positional shifts indicative of implicit communication. (b) \textit{Top:} A diagram of dolphin foraging strategies documented in \protect\cite{neumann2003feeding}. \textit{Middle:} Photos of wolves coordinating while hunting, as shown in \protect\cite{muro2011wolf}. \textit{Bottom:} The learned behavior of our multi-agent system.}
    \label{fig_qual_results}
\end{figure}
\subsection{Animalistic Coordination}
We perform post-hoc qualitative analysis of CD-DDPG trajectories. In the trajectories, the pursuers appear to adjust their position slightly in response to the movements of fellow pursuers as they close in on the evader (Figure \ref{fig_qual_results}a). Moreover, it seems that the pursuers occasionally move away from the evader---something a less coordinated strategy would not do---to maintain the integrity of the group formation. This explains the performance difference between CD-DDPG and the competing analytical strategies, as the potential-field pursuers have no basis for making small-scale adaptive movements. We interpret these results as encouraging evidence that implicit communication has emerged amongst CD-DDPG pursuers.

\subsection{Phantom Coordination}
In Section \ref{sec_results}, we describe ``phantom coordination" as independent action that is falsely perceived as coordination from the perspective of IC. Phantom coordination appears in the greedy pursuit strategy, where we see the IC score for greedy pursuers increase slightly as the velocity of the pursuers decreases. This is counter-intuitive because each greedy pursuer ignores the behavior of its teammates. We would expect the IC score for greedy pursuers to remain flat, mirroring the IC score of the random pursuers.

\begin{figure}[]
    \centering
    \makebox[\linewidth][c]{\includegraphics[width=0.95\linewidth]{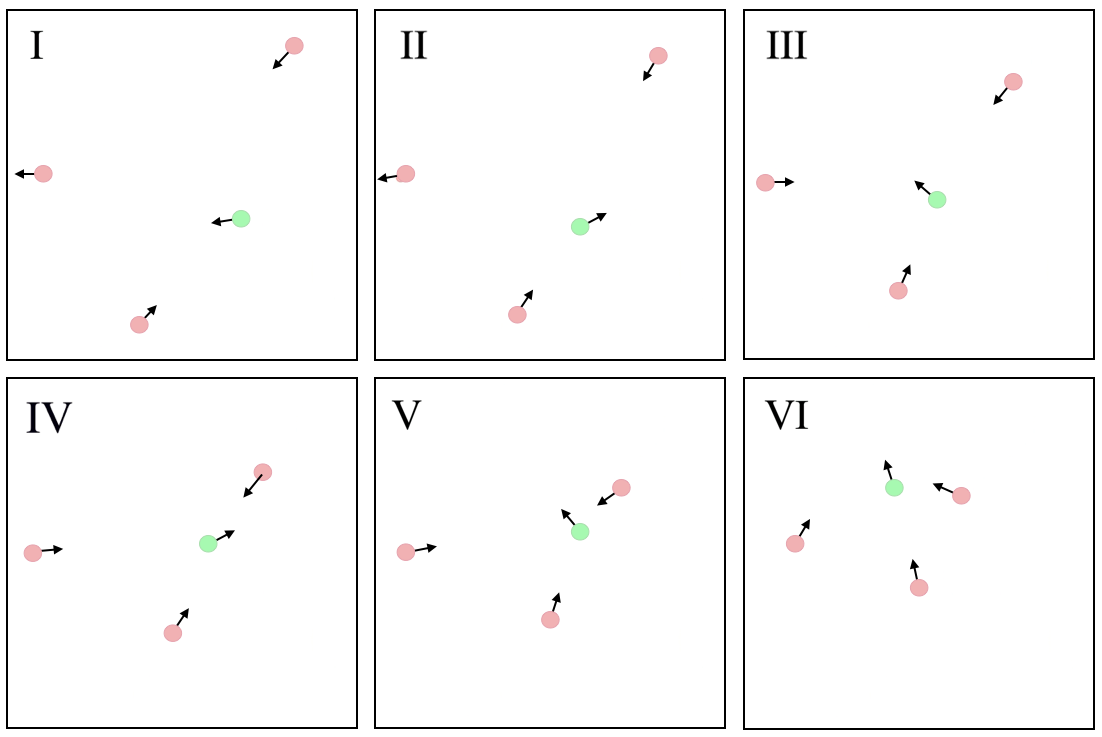}}
    \caption{Snapshots from a 60 time-step trajectory in which phantom coordination appears. Though the pursuers are following independent greedy strategies, their actions produce a triangular formation that is perceived as coordination by the IC performance measure.}
    \label{fig_phantom}
\end{figure}

To diagnose phantom coordination in our environment, we perform qualitative analysis of greedy pursuit at low velocities. In particular, we examine $n=3$ pursuers as they chase an evader at a speed of $\lvert \Vec{v}_p \rvert / \lvert \Vec{v}_e \rvert = 0.4$. The pursuers have no chance of successfully capturing the evader, as evidenced by their capture success performance at this velocity in Figure \ref{fig_cap_success}. However, we find that the straight-line chase patterns of greedy pursuers form temporary triangular patterns around the evader. In Figure \ref{fig_phantom}, the greedy pursuers form an ad-hoc triangular formation around the evader for a duration of 60 time-steps. The leftmost pursuer lies equidistant from the evader around the periodic boundaries and iterates between moving leftward and rightward, depending on which direction creates a shorter line to the evader. The other two pursuers approach the evader from above and below, respectively. This behavior causes the evader to move in a zig-zag pattern in the center of the triangle until a large opening appears, through which the evader can escape.

This behavior leads to phantom coordination because IC is computed between two consecutive time-steps and averaged over whole trajectories. This means that, in the case of the greedy pursuers, IC scores for independent actions are averaged together with subsets of each trajectory that consist of seemingly highly coordinated behavior.
\label{sec_apdx_qual_phantom}

\section{Experiment details}
\label{apdx_experimental_details}
\subsection{Potential-Field Hyperparameters}
The Greedy potential-field defined by Equation \eqref{eq_attractive_potential} is subject to a single hyperparameter $k_{\textrm{att}}$, which defines the strength of the attractive pull of the agent towards its goal. We set $k_{\textrm{att}}=1.5$ in all of our experiments. 

\subsection{Policy Learning Hyperparameters}
Actors $\mu_\phi$ are trained with two hidden layers of size 128. Critics $Q_\omega$ are trained with three hidden layers of size 128. We use learning rates of $1\textrm{e}^{-4}$ and $1\textrm{e}^{-3}$ for the actor and critic, respectively, and a gradient clip of 0.5. Target networks are updated with Polyak averaging with $\tau=0.001$. We used a buffer $\mathcal{D}$ of length $500000$ and sample batches of size $512$. We used a discount factor $\gamma=0.99$. All values are the result of standard hyperparameter sweeps.

\subsection{Experiments}
The models for all experiments were trained for $50000$ epochs of $500$ steps. We reserve $W=1000$ epochs for warm-up, during which the behavioral policy $\beta_0$ is followed. Pursuer velocity is decayed over $v_{\textrm{decay}}=15000$ epochs. Test-time performance, as in Figures \ref{fig_cap_success}, \ref{fig_ic_results}, and \ref{fig_high_ic} is averaged across 100 independent trajectories with separate random seeds. For the ablation in Figure \ref{fig_ablation}, each method was trained for $7500$ epochs across 10 random seeds. All experiments leveraged an Nvidia GeForce GTX 1070 GPU with 8GB of memory.

\subsection{Assets}
The toroidal pursuit-evasion environment used in this work is an extension of the pursuit-evasion environment introduced by \citet{lowe2017multi}. The original environment is open-sourced on Github under the MIT license. We cited the authors accordingly in the main text. New assets (i.e. the new pursuit-evasion environment) are included in the supplementary material. None of the assets used in this work contain personally identifiable information or offensive content.


\end{document}